\documentclass[twocolumn]{aastex631}
\usepackage{amsmath}
\usepackage{subfigure}  %插入多图时用子图显示的宏包
\usepackage{wasysym}
\usepackage{graphicx}
\usepackage{amssymb}
\usepackage{epstopdf}
\usepackage{mathrsfs}
\usepackage{anyfontsize}
\usepackage{CJK}
\usepackage{url}
\usepackage{color}
\usepackage{lipsum}
\usepackage{diagbox}
\usepackage{textcomp}
\usepackage{gensymb}
\usepackage{booktabs}
\usepackage{threeparttable}
\usepackage[figuresright]{rotating}
\usepackage{multirow}

\newcommand{\ed}[1]{\textbf{\color{black}#1}}
\newcommand{\hamr}{h$\alpha$mr}
\newcommand{\lamr}{l$\alpha$mr}
\newcommand{\hamp}{h$\alpha$mp}
\newcommand{\lamp}{l$\alpha$mp}

\shorttitle{}
\shortauthors{Hu et al.}
\begin{document}
%\linenumbers
\begin{CJK*}{UTF8}{gbsn}

	\title{Angular momentum variation of the Milky Way thick disk: The dependence of chemical abundance and the evidence on inside-out formation scenario}
	\correspondingauthor{Zhengyi Shao}
	\email{zyshao@shao.ac.cn}
		
	\author[0000-0003-1828-5318]{Guozhen Hu (胡国真)}
	\affiliation{Key Laboratory for Research in Galaxies and Cosmology, Shanghai Astronomical Observatory, Chinese Academy of Sciences, 80 Nandan Road, Shanghai 200030, China}
	\affiliation{University of Chinese Academy of Sciences, 19A Yuquan Road, 100049, Beijing, China}
	
	\author[0000-0001-8611-2465]{Zhengyi Shao (邵正义)}
	\affiliation{Key Laboratory for Research in Galaxies and Cosmology, Shanghai Astronomical Observatory, Chinese Academy of Sciences, 80 Nandan Road, Shanghai 200030, China}
    \affiliation{Key Lab for Astrophysics, Shanghai 200234, China}
	
	\author[0000-0002-0880-3380]{Lu Li (李璐)}
	\affiliation{Key Laboratory for Research in Galaxies and Cosmology, Shanghai Astronomical Observatory, Chinese Academy of Sciences, 80 Nandan Road, Shanghai 200030, China}
	\affiliation{University of Chinese Academy of Sciences, 19A Yuquan Road, 100049, Beijing, China}

\begin{abstract}

We investigate the angular momentum of mono-abundance populations (MAPs) of the Milky Way thick disk by using a sample of 26,076 giant stars taken from APOGEE DR17 and Gaia EDR3. The vertical and perpendicular angular momentum components, $L_Z$ and $L_P$, of MAPs in narrow bins have significant variations across the [$\alpha$/M]-[M/H] plane. $L_Z$ and $L_P$ systematically change with [M/H] and [$\alpha$/M] and can be alternatively quantified by the chemical gradients: $d[{\rm M/H}]/dL_Z = 1.2\times 10^{-3} $\,dex\,kpc$^{-1}$\,km$^{-1}$\,s, $d{\rm [M/H]}/dL_P = -5.0\times 10^{-3}$\,dec\,kpc$^{-1}$\,km$^{-1}$\,s, and $d[\alpha/{\rm M}]/dL_Z = -3.0\times 10^{-4} $\,dex\,kpc$^{-1}$\,km$^{-1}$\,s, $d[\alpha/{\rm M}]/dL_P = 1.2\times 10^{-3}$\,dec\,kpc$^{-1}$\,km$^{-1}$\,s. These correlations can also be explained as the chemical-dependence of the spatial distribution shape of MAPs. We also exhibit the corresponding age dependence of angular momentum components. Under the assumption that the guiding radius ($R_g$) is proportional to $L_Z$, it provides direct observational evidence of the inside-out structure formation scenario of the thick disk, with  $dR_g/dAge = -1.9$\,kpc\,Gyr$^{-1}$. The progressive changes in the disk thickness can be explained by the upside-down formation or/and the consequent kinematical heating.

\end{abstract}
	
\keywords{Galaxy: disk --- Galaxy: evolution --- Galaxy: kinematics and dynamics --- Galaxy: structure --- methods:
		data analysis --- stars: abundances}

\section{Introduction}\label{sec:intro}

It is widely accepted that the Milky Way disk consists of two prominent components \citep{1983MNRAS.202.1025G}, the dynamically hot thick disk and the cold thin disk (e.g., \citealt{2014A&A...567A...5R,2015A&A...583A..91G,2016MNRAS.461.4246W}). In general, the correspondences between stellar ages and spatial distributions (in both radial and vertical senses) indicate that the overall formation of the disk occurred in an inside-out and upside-down manner from the thick disk to the thin disk \citep{2012ApJ...753..148B,2013ApJ...773...43B,2016ApJ...823...30B,2017ApJ...849...17F,2019ApJ...884...99F}. In this scenario, the older stars formed in an early chaotic bursty mode, with a geometrically thicker layer and a smaller radial scale length, while the younger populations formed in a thinner but radially larger disk. \\

In focusing on the thick disk, the short formation time scale makes it difficult to investigate how its structure formed. Many simulations proposed that the thick disk formed in an inside-out fashion (e.g., \citealt{2003A&A...399..961S,2017MNRAS.467.1154S,2018MNRAS.473..867K}). \ed{But this formation scenario lacks direct and significant observational evidence.}  
For instance, by using G-type dwarfs from SDSS/SEGUE, \cite{2012ApJ...753..148B} found that the scale lengths of the $\alpha$-enhanced sub-populations (dominated by the thick disk) are all similar, suggesting no inside-out signals. \cite{2018A&A...618A..78H} claimed three hints against the inside-out scenario, including that the disk scale length does not increase with time, the radial metallicity gradient of the thick disk is flat, and the chemical abundances have very small dispersions at a given age. Recently, \cite{2021A&A...655A.111K} investigated Apogee DR16 stars in the [$\alpha$/Fe]-[Fe/H] plane and found that the ridge-lines of sub-samples with different guiding radii slightly change with the [$\alpha$/Fe], which can be regarded as weak evidence of the inside-out formation of the thick disk. \\

There are two crucial observational aspects in investigating the thick disk structure formation, the stellar initial spatial distribution and the stellar age. Unfortunately, both of these are difficult to obtain. \\

First, it is a challenge to deduce the spatial distribution of stars at birth from their present-day Galactic radius due to the radial migration ($blurring$ and $churning$), which can redistribute the stars and ambiguate the memory of their initial kinematic status \citep{2012ApJ...754..124Y,2014ApJ...794..173V,2014ApJ...788...89T,2018ApJ...863...93J}. \textbf{$Blurring$} is a process in that disk stars move away from their born positions through epicyclic motion \citep{2009MNRAS.396..203S,2014RvMP...86....1S, 2009A&A...501..941H,2009MNRAS.397.1286A,2019MNRAS.489..176M}. It conserves the angular momentum of individual stars. \textbf{The} $churning$ process may occur through the resonant interaction between stars and the non-axisymmetric structures of the gravitational potential (spirals or bar) \citep{2002MNRAS.336..785S,2012MNRAS.422.1363S,2012MNRAS.426.2089R}. It changes the angular momentum of individual stars, but for a single population of a given birth radius, this process will increase the dispersion in angular momentum while roughly keeping the average value (see Figure 3 of \citealt{2016MNRAS.457.2107S}) and does not "heat" the disk radially \citep{2015A&A...580A.126K}.　\ed{Therefore, using the angular momentum ($\textbf {\emph L}$) of the stars  instead of their Galactic radius may greatly mitigate the problems caused by migration.}  \\

Additionally, the angular momentum is the most fundamental quantity of disk galaxies. It is a better indicator of the disk formation and the gas-infall process \citep{2018ApJ...868..133F}. The vertical angular momentum component $L_Z$ can characterize the radial sense of the disk, and the perpendicular component $L_P$ relates to the thickness of the disk. Overall, the variation of angular momentum of disk stellar populations is essential in recovering the thick disk structure formation process. \\

\textbf{Another difficulty is in acquiring accurate stellar age. Current age measurements of individual stars have large uncertainties (\citealt{2016ApJ...823..114N}) comparable to the short formation timescale of the thick disk, which makes it difficult to intuitively describe the star formation history of the thick disk. 
%Instead, many studies have used chemical abundance as a proxy (e.g. \citealt{ 2012ApJ...753..148B,2016ApJ...823...30B,2013A&A...560A.109H,2017ApJ...834...27M,2019MNRAS.489.1742F}). That is because the gas-infall and star formation processes during the disk formation are closely related to chemical enrichment. Therefore, tracing the chemical evolution can better describe the thick disk formation.} 
Alternatively, because the gas-infall and star formation processes during the disk structure formation are closely related to chemical enrichment, many studies have used chemical abundance as a proxy of stellar age (e.g. \citealt{ 2012ApJ...753..148B,2016ApJ...823...30B,2013A&A...560A.109H,2017ApJ...834...27M,2019MNRAS.489.1742F}). }
\\

Recent works \textbf{suggested} employing mono-abundance populations (MAPs). A MAP consists of stars with similar abundance, e.g., metallicity [M/H] and the $\alpha$-enhancement [$\alpha$/M] (e.g.,  \citealt{2012ApJ...751..131B,2016ApJ...823...30B,2020MNRAS.492.3631M}), which are supposed to have a common origin with similar ages and chemical enrichment histories. Usually, MAPs are regarded as mono-age populations, making them physically intuitive indicators of galaxy formation processes (e.g., \citealt{2012ApJ...751..131B,2016ApJ...823...30B,2017ApJ...834...27M}). \\

Therefore, it is worthwhile to investigate the variation of the angular momentum of different MAPs of the thick disk to reveal its structure formation. This paper presents such an investigation based on a large sample of giant stars (\citealt{2022ApJ...929...33H}, hereafter HS22) from APOGEE DR17 (\citealt{2022ApJS..259...35A}) and Gaia EDR3 \citep{2021A&A...649A...1G, 2021A&A...649A...2L}. We analyze the distribution of orbital parameters of small MAPs of the thick disk, including the angular momentum components and orbital eccentricity, to detect the variation and evolution of these parameters and subsequently discuss their dependence on chemical abundance. \\

Additionally, although the current stellar age measurements have large uncertainties for individual stars, it is still possible for us to statistically constrain the typical \textbf{age of a given MAP}. Then we can directly obtain the correlation between stellar age and angular momentum and recreate the thick disk's radial and vertical structure formation scenario. \\

This paper is organized as follows. In the next section, we introduce the thick disk sample and the observational or derived parameters of sample stars. Section~\ref{sec:Properties} shows the results of orbital parameters of MAPs, including the angular momentum components and orbital eccentricity. In Section~\ref{sec:discussion}, we first discuss the correlations between angular momentum and chemical abundances and then use the \cite{2018MNRAS.481.4093S} catalog of stellar age to debate the dependence of angular momentum with stellar age and finally generalize these relationships to a disk formation scenario. Conclusions are summarized in Section~\ref{sec:Conclusion}. \\   

\section{Sample and Data}\label{sec:data}

\subsection{Thick disk sample}\label{subsec:Sample}
 
In HS22, we have identified and quantified four sub-disk components of the Milky Way by using the chemical abundances ([M/H], [$\alpha$/M]) and the 3D velocities ($ V_{R}, V_{\phi}, V_{Z}$)  taken or derived from APOGEE DR17 and Gaia EDR3 for a sample of 119,558 giant stars. We briefly describe the procedure of HS22 below. Firstly, the sample stars were divided into [M/H] bins. For each bin, we used the multivariate Gaussian Mixture model (GMM) to fit the subsample with other four variables, [$\alpha$/M], $V_{R}$, $V_{\phi}$ and $V_{Z}$. In the range of $-1.2<$[M/H]$<0.4$\,dex, all bins were well modeled by two Gaussian components belonging to the high-$\alpha$ or low-$\alpha$ sequences separately. These two sequences were further separated \textbf{at} [M/H]$=-0.1$\,dex to denote metal-poor or metal-rich parts. Thus, four sub-disk components were determined and named as \hamp, \hamr, \lamp, and \lamr\, separately. Of these, the \hamp\, component, which corresponds to the canonical thick disk, has the hottest kinematics and the smallest radial distribution (see Fig. 2 and 5 of HS22). It is significantly distinguished from the other three sub-disks. In this paper, we follow the work of HS22 and \textbf{select} only the thick disk (the \hamp\, component) for investigation. \\

\begin{figure}[htbp] 
	\centering
	\includegraphics[width=0.5\textwidth]{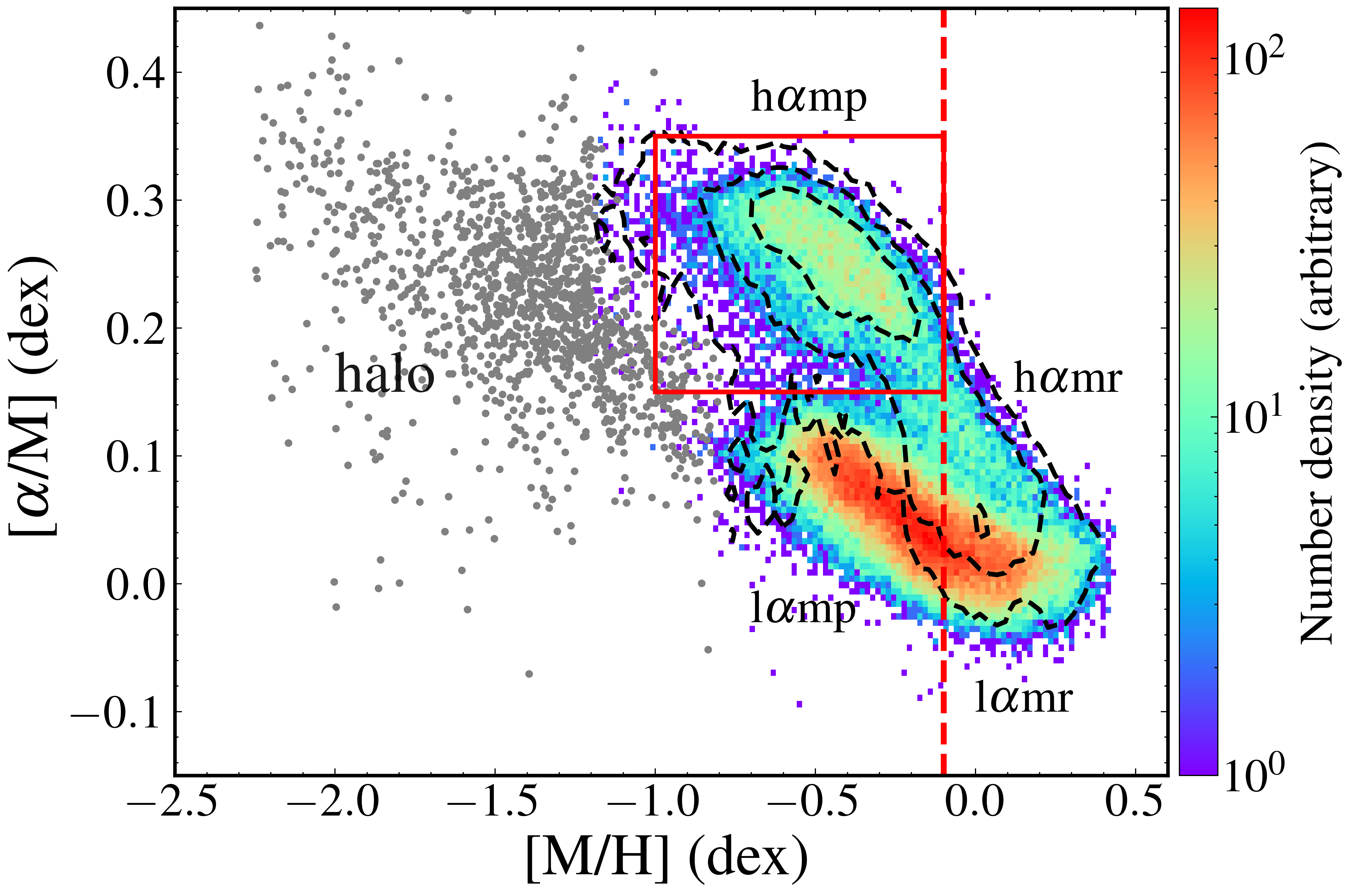} 
	\caption{Number density distribution of HS22 sample stars in the [$\alpha$/M]-[M/H] plane. The colored distribution is for the disk components and black-dashed lines represent the 1 and 2 $\sigma$ contours for the high-$\alpha$ sequence only. The red dashed vertical line at [M/H]=-0.1\,dex dividing the metal-poor and metal-rich regions.  The red box delineates the canonical thick disk (\hamp\,) region selected for this paper. Halo stars are shown as grey dots.}
	\label{Fig:1}
\end{figure}

We restrict the thick disk sample stars in the region of $-1.0 < $ [M/H] $ < -0.1$\,dex and $0.15 <$ [$\alpha$/M] $< 0.35$\,dex (the red box in Fig.~\ref{Fig:1}). In this region, there are 26,076 giant stars having good quality of chemical and astrometrical data from APOGEE DR17 and Gaia EDR3. As a by-product of the GMM, we have calculated the membership probabilities for individual stars belonging to each $\alpha$-sequence (Table 2 of HS22). In this work, we select 24,935 thick disk stars ($\sim 96 \%$ stars within the given region) having membership probability $P_{h\alpha} > 0.99$. This choice ensures us that the thick disk sample has very little contamination from other sub-disk components. \\  

\subsection{Angular momentum and orbital eccentricity}\label{subsec:orbit}

In HS22, for each sample star, we have calculated the 6D kinematical information (3D positions and 3D velocities) relative to the Sun, by using the coordinates and proper motions from Gaia EDR3, the radial velocities from APOGEE DR17 and the photo-geometric distances estimated by \cite{2021AJ....161..147B}. They have been transformed into the Galactocentric cylindrical system $(R_{\rm gc}, \phi, Z, V_{R}, V_{\phi}, V_{Z})$ by assuming that the Galactocentric distance of the Sun is $R_\odot= 8.125$ kpc \citep{2018A&A...615L..15G} and its height from the Galactic plane is $Z_\odot=20.8$ pc \citep{2019MNRAS.482.1417B}, and the solar motions are (11.1, 242, 7.25) km\,s$^{-1}$ in the radial, rotational and vertical directions, respectively (\citealt{2010MNRAS.403.1829S,2012ApJ...759..131B}). In this work, we subsequently use the 6D kinematic data as the input of the GALPOT package to integrate the stellar orbit of each sample star under the Galactic potential model of \cite{2017MNRAS.465...76M}. Thus, we obtain the vertical and perpendicular angular momentum components $L_Z$ and $L_P$. Additionally, we obtain the Galactic pericenter ($r_{\rm peri}$) and apocenter ($r_{\rm apo}$) radii, and then derive the orbital eccentricity defined as $ecc = (r_{\rm apo}-r_{\rm peri})/(r_{\rm apo} + r_{\rm peri})$. \\

Usually, the $L_Z$ of star can be transferred to its guiding radius ($R_g$) based on the relationship $V_{\phi}^2$/$R_{g}$=$L_{Z}^2$/$R_{g}^3$ \citep{2014ApJ...788...89T}. We calculate the $R_g$ for each star under the assumption that the rotational velocity is a constant across the region covering our thick disk sample, with $V_{\phi} = 230$\,km\,s$^{-1}$ (\citealt{2010MNRAS.403.1829S,2012ApJ...759..131B}).  These value-added measurements ($L_Z$, $L_P$, $ecc$ and $R_g$) are listed in Table~\ref{tab:1} and its machine-readable form is available.\\

\begin{sidewaystable}[h] 
\setlength{\abovecaptionskip}{0.05cm} 
\centering
\caption{Summary of parameters for thick disk stars} \label{tab:1}
\begin{threeparttable}
\begin{tabular*}{\hsize}{@{\extracolsep{\fill}}llrrrrrccrrr} 
\hline
\hline

APOGEE-ID~\tnote{a}& SOURCE-ID~\tnote{b} & {[$\alpha$/M]}~\tnote{c}& {[M/H]}~\tnote{d} &  Age ~\tnote{e} &  $P_{h\alpha}$ ~\tnote{f}& $L_{Z}$ & $L_{P}$ & $ecc$ &$R_g$& $R_{gc}$\\
 & & dex & dex & Gyr &  & kpc km\,s$^{-1}$ & kpc km\,s$^{-1}$  & & kpc &kpc \\
\hline

  2M00080293+1354102 &   2768636008919583488 &   0.155 &    -0.902 &  ...  &     0.62  &     2271.86   &  588.26&      0.14 & 9.00 &9.89 \\
  2M00095633+6739200 &   528907737695674496 &    0.202 &    -0.929 &  4.61 &    0.45  &    558.74 &    442.44 &    0.78  &10.24& 2.43\\
  2M03544189-6938299 &  4666648650490972800 &   0.199  &    -0.928&   9.00 &    0.80&      1033.97&    231.20&     0.42&7.89&4.50\\  
  2M05200684+2919227 &  3446462494832460672 &  0.207 &    -0.925&    6.43  &    0.84&     1457.72&    434.60&      0.48&11.57&6.34\\
  2M09474458+0244510  & 3847051689345915648  & 0.226 &     -0.469 &   8.56&    1.00   &   409.23&    371.37 &      0.83&9.72&1.78\\   
  2M09482299+4600232 &  821455518049745280 &   0.207 &     -0.498 &   7.44&    1.00   &   1394.64&     979.77&        0.38&9.77&6.06\\
  2M09525110+3132524 &  745175382746994432 &   0.230 &     -0.455 &   ...    &  1.00  &    1368.62 &    232.65&       0.35&8.93&5.95\\  
  2M09530067+3752199 &  799998372538133120 &   0.267 &   -0.465 &   7.41 &     1.00  &       1801.05&     370.35&        0.39&9.64&7.83\\ 
 2M18341526-1304434 &  4104893557043564160 &  0.191&    -0.406&    2.94&   0.96&      890.77&     91.96&      0.18&3.45&3.89\\   
  2M18344194-2914203 &  4048233180241935744 &  0.221&     -0.458&  4.72&    1.00 &  155.23&    231.57&     0.62&2.37&0.67\\

\hline
\hline
\end{tabular*}
\tablecomments{The complete table is available on the publisher website.}
\begin{tablenotes}
	\footnotesize
	\item[a] APOGEE DR17 object name.
    \item[b] Gaia EDR3 source id.
	\item[c] [$\alpha$/M] in APOGEE DR17.
	\item[d] [M/H] in APOGEE DR17.	
	\item[e] \citet{2018MNRAS.481.4093S}. 
	\item[f] Table 2 of \citet{2022ApJ...929...33H}. The membership probabilities belonging to high-$\alpha$ for individual sample stars. 
\end{tablenotes}
\end{threeparttable}
\end{sidewaystable}
%--

\subsection{Stellar age}\label{subsec:age}

\cite{2018MNRAS.481.4093S} provided a catalog of stellar parameters, including the stellar age, based on the Bayesian framework inference using broad-band photometric, spectroscopic and astrometric information. This catalog matches our giant star sample to a large extent. \ed{We rejected stars with age uncertainties $\sigma_\tau > 1.5$\,Gyr. Our final sample contains 11,761 giant stars having age measurements, which is about 47\% of the sample.}\\

It should be mentioned that \cite{2018MNRAS.481.4093S} use a prior on age, metallicity, and position in their Bayesian framework, i.e., stars of the thick disk component tend to be older and more metal-poor than stars of the thin disk component. So the prior is not fully independent between stellar age and metallicity, though it has a very wide distribution on the [Fe/H]-age plane. However, we \textbf{assume} that the prior of \cite{2018MNRAS.481.4093S} is reasonable for the MW components, so the age measurements for individual stars are reliable. They are independent to the angular momentum and should be suitable in subsequent discussion of the angular momentum evolution (cite Sec.~\ref{subsec:AgeDependence}). \\

\section{Angular momentum and orbital Properties across [$\alpha$/M]-[M/H] plane} \label{sec:Properties}

We split the stellar sample into mono-abundance bins with bin size of 0.025 $\times$ 0.025 dex$^2$ for [M/H] and [$\alpha$/M]. This bin size is chosen to be slightly larger than the typical abundance uncertainties of the Apogee catalog (0.01 $\sim$ 0.02 dex), but it is small enough to resolve the thick disk region (the red box in the Fig.~\ref{Fig:1}) in the [$\alpha$/M]-[M/H] plane. \\

\begin{figure*}[htbp] 
	\centering
	\includegraphics[scale=0.6]{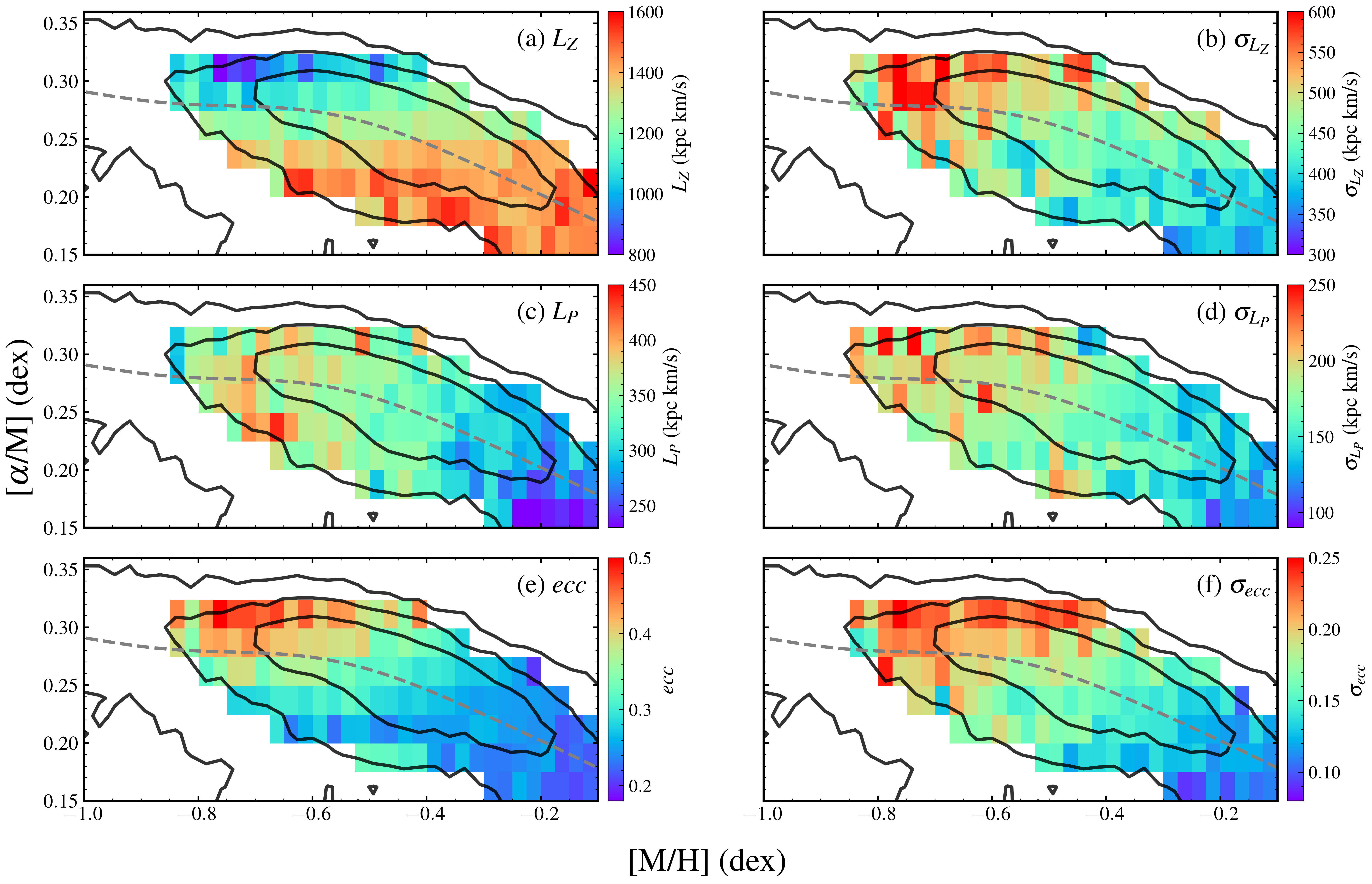} 
	\caption{Distributions of the MAP's parameters in the [$\alpha$/M]-[M/H] plane. Parameters are coded by color, including $L_{Z}$, $L_{P}$, $ecc$ and the dispersions $\sigma_{L_Z}$, $\sigma_{L_P}$ and $\sigma_{ecc}$, respectively. The black lines represent $1\sigma$, $2\sigma$ and $3\sigma$ contours of the number-density distribution of the thick disk sample star. The grey dashed line curves the ridge-line of the number density, which is identified by a robust Gaussian process \citep{2021A&C....3600483L}. }

	\label{fig:2}
\end{figure*}

We select the MAPs with the number of star $n$ $>$ 30 for investigation. These MAPs cover the dominant part of the thick disk region ($\sim 2\sigma$ of the number density contour in Fig.~\ref{fig:2}). For parameters that will be discussed in this paper, such as the angular momentum components ($L_Z, L_P$), the orbital eccentricity ($ecc$) and the stellar age, we notice that they all have non-negligible dispersions for each MAP due to the intrinsic distributions and/or the observational uncertainties. Nevertheless, we suppose that their median values are statistically robust and could be reasonably adopted to determine the variations across the entire thick disk region. Therefore, in the following context, each MAP's properties are specifically referred to their median values unless otherwise stated. The corresponding dispersions ($\sigma$) of parameters for each MAP are quantified in term of half of the 16\% to 84\% interval respectively. \\

In Fig.~\ref{fig:2}, we plot $L_Z, L_P$ and $ecc$ of MAPs on the [$\alpha$/M]-[M/H] plane. These parameters show significant variations and systematic changes across the metallicity and the $\alpha$-enhancement. \\

\begin{description}
  
    \item[Angular momentum] Panels (a) and (c) of Fig.\ref{fig:2} plot the color-coded $L_{Z}$ and $L_P$ values. 
    %There are about twice \textbf {two times??} variations for 
    \ed{Both of these two angular momentum components show significant variations, with $L_Z$ from 800 to 1600 kpc\,km s$^{-1}$ and $L_P$ from 230 to 450 kpc\,km s$^{-1}$. } The $L_{Z}$ increases/decreases with the metallicity/$\alpha$-enhancement and the $L_P$ follows an opposite trend. One can also find that they present monotonous changes along the number-density ridge line (the dotted line in each panel), which represents the well-known chemical abundance evolutionary trajectory (e.g. \citealp{2019A&A...623A..60S}, see their Fig.2; \citealp{2020MNRAS.497.2371L}, see their Fig.7). Notably, \ed{the distribution of $L_Z$} appears to have a horizontal structure in the [$\alpha/M$]-[M/h] plane, which means that, at a given metallicity, higher-$\alpha$ MAPs tend to have lower $L_Z$ values.    \\
    
    Panels (b) and (d) show the corresponding angular momentum dispersions, $\sigma_{L_Z}$ and $\sigma_{L_P}$. \textbf{Both of them }decrease with the metallicity, indicating that MAPs at more metal-poor stage have larger angular momentum dispersions. \textbf{This phenomenon could be} mainly due to the $churning$ effect, which disperses the angular momentum distributions for a given stellar population. \\

    \item[Orbital eccentricity] It shows a systematic variation \ed{of $ecc$} along the evolutionary trajectory, from 0.50 to 0.18, with more metal-poor stars having larger eccentricities (panel (e)). Interestingly, \ed{the distribution of $ecc$} also has a horizontal structure, which is similar to that of the $L_Z$ in panel (a) but with the opposite trends along the evolutionary trajectory. It implies an anti-correlation between $L_Z$ and $ecc$ of the current sample. Notably, this anti-correlation is partly attributed to the effect of the $blurring$ radial migration (e.g., \citealt{2014ApJ...788...89T,2016MNRAS.462.1697A}). Disk stars with larger orbital eccentricities are more probably to be observed around their apocenter points rather than near the pericenter points. So apparently, a solar-centric sample may contain more inner (less $L_Z$ or $R_g$) stars with more eccentric orbits. This kind of observational selection bias actually helps us to extend our sample coverage towards the inner part of the MW disk. \\
    
    Panel (f) shows the corresponding dispersion of orbital eccentricity, \ed{where $\sigma_{ecc}$ decreases} with the metallicity, indicating that more metal-poor MAPs have larger eccentricity dispersions, which maybe also due to the $blurring$ effect for older stellar populations.  \\
    
\end{description}

In summary, we find that the thick disk MAPs have significant variations of angular momentum and orbital eccentricity across the [$\alpha$/M]-[M/H] plane, which implies \textbf{a} strong correlation of its formation scenario and chemical evolution.   \\

\section{Discussions} \label{sec:discussion}

\subsection{Chemical dependence of angular Momentum}\label{subsec:ChemicalDependence}

\begin{figure*}[htbp] 
	\centering
	\includegraphics[width=1\textwidth]{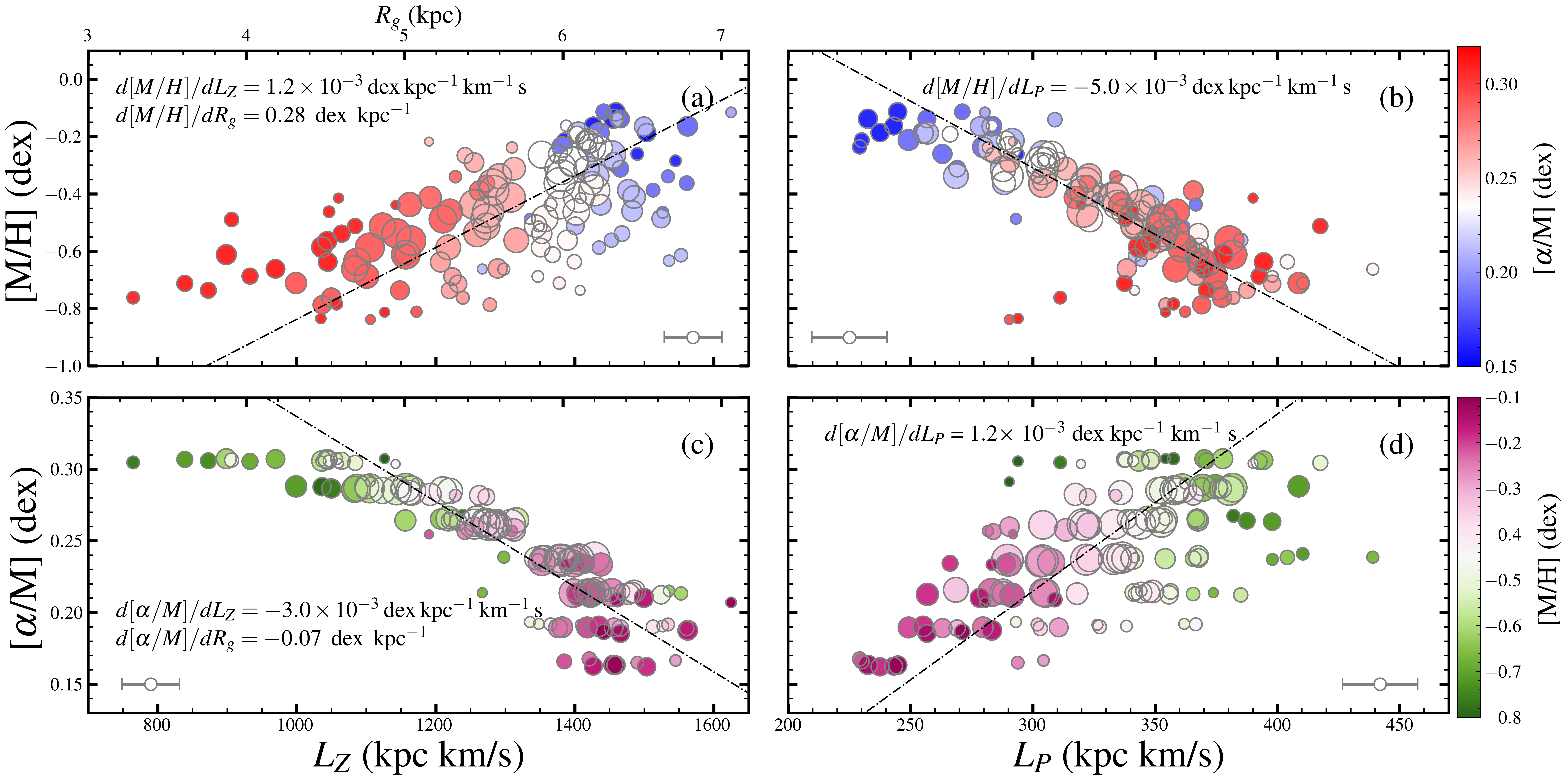} 
	\caption{Correlations between chemical abundance and angular momentum of MAPs. Top panels: for the [M/H]; Bottom panels: for the [$\alpha$/M]. Different angular momentum components, $L_{Z}$ (or $R_g$) and $L_{P}$ are plotted on the left and right panels, respectively. The symbol size represents the star number $n$ of MAP, from 33 to 428. The symbol color are coded by the median values of [$\alpha$/M] for the top panels, and [M/H] for the bottom panels. The error bars in the corner of each panel represent the typical uncertainties of the median values of corresponding parameters, which is estimated as $\sigma/\sqrt{n}$. The slopes of each correlation (dash-dotted lines) are fitted linearly with the uncertainties of median value to be taken into account. }
	\label{fig:3}
\end{figure*}

\ed{If we treat the MAPs as independent observational data points, we can directly detect the correlations between chemical abundances ([M/H] or [$\alpha$/M]) and angular momentum components ($L_Z$ or $L_P$). Since $L_Z$ or $L_P$ is closely related to the radial scale or thickness of the disk, these two correlation slopes can be regarded as substitutes for the radial and vertical gradients of chemical abundances.}\\

\subsubsection{$L_Z$ and radial gradients of chemical abundances}\label{subsec:radial-gradients}

The correlation between [M/H] and $L_Z$ is \textbf{visible} in panel (a) of Fig.~\ref{fig:3}, with the slope of $d[{\rm M/H}]/dL_Z = 1.2\times 10^{-3} $\,dex\,kpc$^{-1}$\,km$^{-1}$\,s, or $d[{\rm M/H}]/dR_g = 0.28 $\,dex\,kpc$^{-1}$. \ed{It indicates the inverse $L_Z$ or radial gradients of [M/H], suggesting that the stars born in the outer part of the thick disk, which exhibit larger $L_Z$ (or $R_g$), are more metal-rich than those in the inner part.} 
This correlation is opposite to the overall radial gradient of the MW disk, which declares that the outer (thin) disk is more metal-poor than the inner (thick) disk, but it supports the model predictions of chemical radial gradient for the individually considered thick disk \citep{2017MNRAS.467.1154S}. \\

We notice that there is a large scatter of the [M/H]-$L_Z$ correlation, which is mainly due to the different [$\alpha$/M] values. For MAPs with a given [M/H], e.g. [M/H]$\sim-0.5$, the [$\alpha$/M] significantly decrease with $L_Z$ (or $R_g$) (shown as the changes of color in panel (a) of Fig.~\ref{fig:3}). These MAPs have the same [M/H] but different $\alpha$-enhancement, which implies that they have different chemical-\ed{enrichment} processes. \ed{It agrees that, in the inner region of the thick disk, stars formed in a denser gas environment may have a higher star formation rate leading to a more rapid $\alpha$-enhancement \citep[e.g.,][]{2016A&A...589A..66H}.}   \\

The $L_Z$ dependence of [$\alpha$/M] is also shown as a tight relationship between [$\alpha$/M] and $L_Z$ (or $R_g$), where the smaller $L_Z$ (or $R_g$) MAPs have larger [$\alpha$/M] values (panel (c) of Fig.~\ref{fig:3}).  Despite the [$\alpha$/M]-$L_Z$ correlation \ed{is not linear in this work}, we estimate the overall slope of $d[\alpha/{\rm M}]/dL_Z = -3.0\times 10^{-4} $\,dex\,kpc$^{-1}$\,km$^{-1}$\,s, alternatively $d[\alpha/{\rm M}]/dR_g = -0.07$\,dex\,kpc$^{-1}$, which claims a significant $L_Z$ or radial gradient of [$\alpha$/M]. 
\ed{The [$\alpha$/M]-$L_Z$ ($R_g$) correlation is consistent with the findings of \cite{2021A&A...655A.111K}, who analyzed the giant stars of the thick disk from APOGEE DR16. Their work compared the ridge lines in the [$\alpha$/Fe]-[Fe/H] plane for sub-samples with varying $R_g$ and found the inner stars having higher [$\alpha$/Fe] positions.}\\

The radial gradients of chemical abundances of the thick disk have been explored in many previous works \citep[e.g.,][]{2012AJ....144..185C,2014A&A...572A..33M,2018ApJ...860...53L}. Most of them reported weak or none gradients. Probably, that is because they have only use the current Galactocentric distance $R_{gc}$ of the sample star while the radial migration throughout the disk may eliminate the radial metallicity gradient \citep{2018ApJ...860...53L}. In this work, we employ the guiding radius ($R_g$), or the vertical angular momentum ($L_Z$), which can avoid the $blurring$ migration effect for individual stars and also make the $churning$ effect negligible for the MAPs. So, the MAPs' average properties \ed{describe} the original condition of thick disk at most and the initial radial gradients can be more easily exhibited. \\

\subsubsection{$L_P$ and vertical gradients of chemical abundances}\label{subsec:vertical-gradient}

Panels (b) and (d) of Fig.~\ref{fig:3} show the significant chemical-$L_P$ correlations of MAPs. \ed{Firstly,} there is a $L_P$ gradient of metallicity with $d[M/H]/dL_P = -5.0\times 10^{-3}$\,dex\,kpc$^{-1}$\,km$^{-1}$\,s. Considering the $L_P$ uncertainties are quite large (typically shown as the error bar in panel (b)), the [M/H]-$L_P$ correlation should be extremely tight with very small internal scatters. \ed{Secondly}, the correlation between [$\alpha$/M] and $L_P$ is also significant (in panel (d)) but with a much larger scatter than the typical uncertainty of $L_P$. The slope is $[\alpha/{\rm M}]/dL_P = 1.2\times10^{-3}$\,dex\,kpc$^{-1}$\,km$^{-1}$, indicating an inverse $L_p$ gradient of [$\alpha$/M]. Moreover, we can find that the scatter is mainly due to the different [M/H] values, which can be regarded as the conditional  [M/H]-$L_P$ correlation at given [$\alpha$/M]. Therefore, we conclude that the [M/H]-$L_P$ correlation dominates the relationship among these three parameters. \\

Since the $L_P$ value decides the height from the Galactic plane of a star can reach, $Z_{\rm max}$, the chemical dependencies on $L_P$ then correspond to the vertical gradients of chemical abundance. \ed{That means, for our thick disk sample, the thinner distributed populations (with smaller $L_P$) have more metal-richness and less $\alpha$-enhancement}. Previous works reported weak vertical gradients when the current height value $|Z|$ is used \ed{as diagnostics} \citep[e.g.,][]{2018ApJ...860...53L,2019ApJ...880...36Y}. \ed{We suggest} that the $L_P$ gradients of chemical abundances are intrinsic features of the thick disk due to its formation, and the use of $|Z|$ weakens this phenomenon.\\

Moreover, if MAPs with larger $L_P$ values are expected to have larger vertical velocity dispersions ($\sigma_{V_Z}$), the chemical dependence of $L_P$ will also be reflected as the chemical-$\sigma_{V_Z}$ relationships, which has been already shown in previous works. For example, HS22 reported that, $\sigma_{V_Z}$ of the thick disk (the \hamp\, disk) monotonously decreases with [M/H] from 71 to 31 kms$^{-1}$ by using the same giant star sample of this paper. \cite{2013A&A...560A.109H} found that $\sigma_{V_Z}$ progressively decreases with the decreasing of [$\alpha$/Fe] from about 50 to 25 km s$^{-1}$ for a sample of thick disk dwarf stars (see their Fig.11), suggesting that the star formation proceeded in progressively thinner layers. \\

\subsection{Age dependent angular momentum}\label{subsec:AgeDependence}

\begin{figure*}[htbp] 
	\centering
	\includegraphics[scale=0.6]{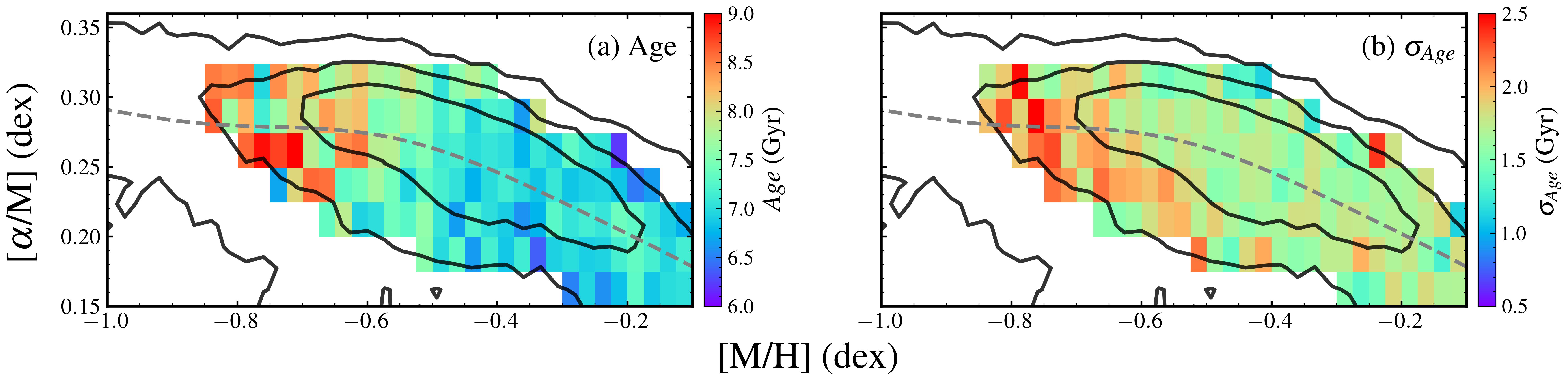} 
	\caption{Distributions of the MAP's stellar age (panel (a)) and the dispersions $\sigma_{Age}$ (panel (b)) in the [$\alpha$/M]-[M/H] plane. The contour and dash line are the same as in Fig.\ref{fig:2}. }

	\label{fig:4}
\end{figure*}

The star-forming process \textbf{leads} to a sustained metallicity enrichment, accompanied by a decrease in the $\alpha$-enhancement. Therefore, the change of chemical abundance in forming stellar populations is usually regarded as an indication of time. So the corresponding changes of the angular momentum components with the chemical abundance can be considered as the structure evolution \citep{2012ApJ...753..148B,2013ApJ...773...43B,2016ApJ...823...30B}. Nevertheless, the direct correlation between stellar age and angular momentum is still expected to explore the structure formation process intuitively.\\

We employ the stellar age measurements of the \cite{2018MNRAS.481.4093S} catalog, and plot the age of MAPs across the [$\alpha$/M]-[M/H] plane in Fig.\ref{fig:4}. The median ages of MAPs vary from 6.0 to 9.0 Gyr (panel (a)). The $\sigma_{Age}$ of MAPs are slightly different, with older MAPs having larger age dispersion (panel (b)). If we estimate the uncertainty of the median value as $\sigma_{Age} / \sqrt{n}$, then the errors are from 0.11 to 0.60 Gyr, with a typical value of $\sim$0.25 Gyr. \ed{The age variation of MAP follows the trajectory of chemical enrichment. More metal-poor or more high-α MAPs exhibiting larger stellar ages, which reflects that the age dependence of chemical abundances of the thick disk is substantial.} It reveals the process of chemical evolution, though the thick disk is believed to have a \textbf{bursty} star formation. This result confirms previous works, which point out the tight age-[M/H] or age-[$\alpha$/M] correlations of the thick disk (e.g., \citealt{2013A&A...560A.109H,2017ApJ...834...27M}).\\

\begin{figure}[htbp] 
	\centering
	\includegraphics[width=0.5\textwidth]{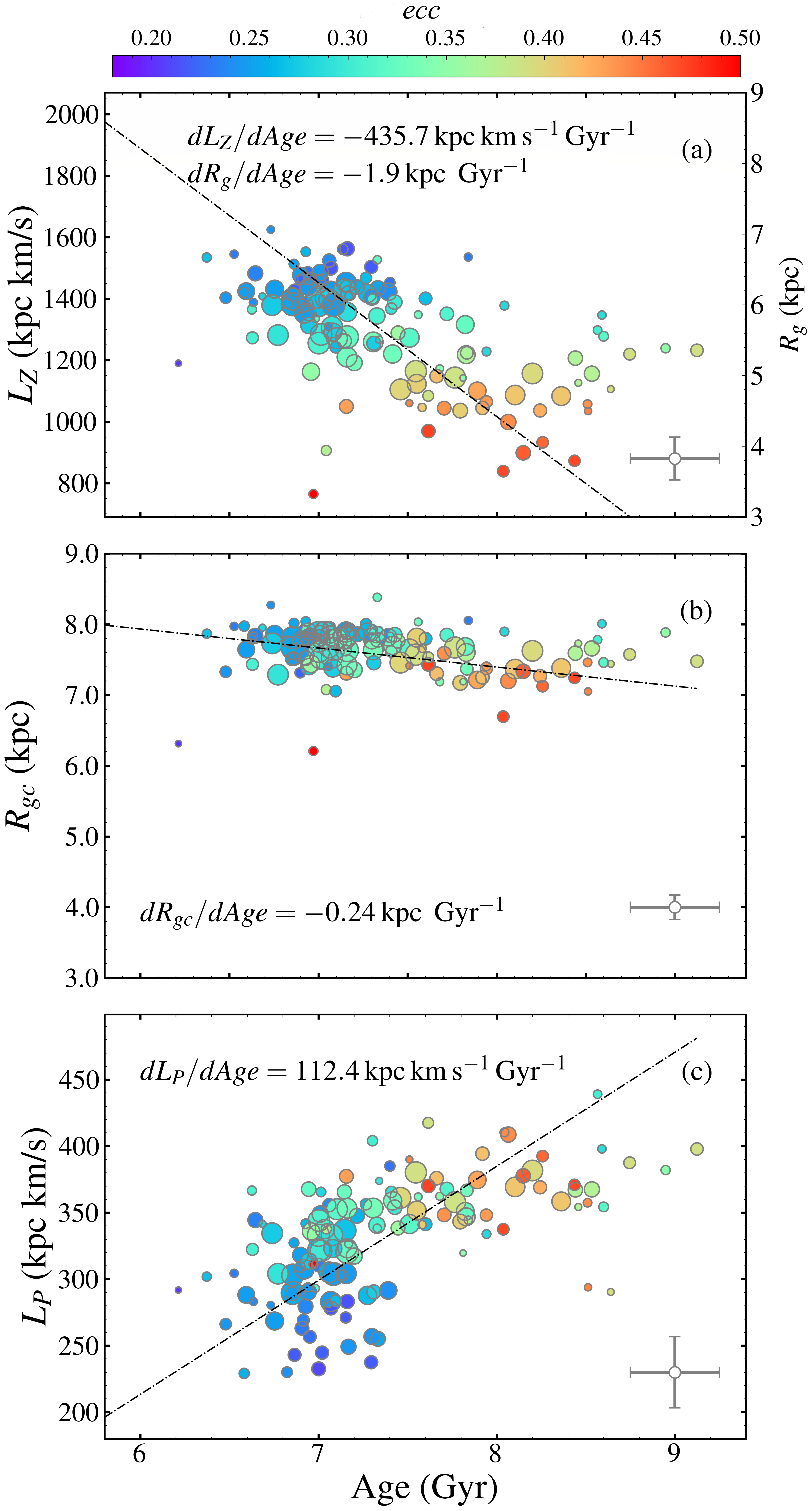} 
	\caption{Correlations between MAP's parameters and stellar age. $L_{Z}$ (or $R_g$), $R_{gc}$ and $L_{P}$ are plotted in separate panels, with each symbol representing a MAP. The color is coded by the median value of $ecc$. The symbols with error bars in the corners of each panel represent the typical uncertainties of the median values of corresponding parameters. The slopes of each correlation (dash-dotted lines) are fitted using the orthogonal distance regression, and the uncertainties of the median values are taken into account. ({\url{https://docs.scipy.org/doc/scipy/reference/odr.html)}}}
	\label{fig:2_par_age}
\end{figure}

Similarly, by using MAPs as \ed{observational data points}, we can directly investigate the correlation between the age and angular momentum. As shown in \textbf{panel} (a) of Fig.~\ref{fig:2_par_age}, the $L_Z$ of MAP decreases substantially with stellar age, having $dL_Z/dAge = -435.7$\,kpc\,km\,s$^{-1}$\,Gyr$^{-1}$. This slope corresponds to $dR_g/dAge = -1.9$\,kpc\,Gyr$^{-1}$ alternatively. We claim that such a steep slope is strong evidence of the radial enlargement of the thick disk.  \\  

As a comparison, we plot the current Galactic radius $R_{gc}$ $vs$ age in panel (b) \textbf{of Fig.~\ref{fig:2_par_age}}. The correlation is also significant, but the dependence is very weak with $dR_{gc}/dAge = -0.24$\,kpc\,Gyr$^{-1}$. Generally, $R_{gc}$ is larger than $R_g$ for all MAPs with different increments according to their $ecc$ values. It is just the impact of the $blurring$ radial migration. MAPs with  more eccentric orbits can move further than their birthing places ($\sim R_g$) and tend to be observed near their apocentre. That is why many works found no (or only slight) enlargement of the thick disk radial size if they only used the $R_{gc}$ for diagnosing \citep{2012ApJ...753..148B,2016ApJ...823...30B}. In other words, it is difficult to detect the inside-out feature of the thick disk formation because the stars are no longer located where they were born due to the radial migration. \\

Panel (c) \textbf{of Fig.~\ref{fig:2_par_age}} shows the correlation between $L_P$ and age, with younger MAPs having smaller $L_P$ values and $dL_P/dAge = 112.4$\,kpc\,km\,s$^{-1}$\,Gyr$^{-1}$. This correlation exhibits the phenomenon that the spatial distributions of stars in younger MAPs formed in thinner layers, or the older MAPs were getting thicker after star formation. \\

\subsection{Structure formation scenario of the thick disk} \label{subsec:Scenario}

Combining all the angular momentum correlations \ed{discussed in sections~\ref{subsec:ChemicalDependence} and \ref{subsec:AgeDependence}, we have} solid observational evidence for the thick disk formation. Older (or relatively metal-poor/high-$\alpha$) stellar populations have smaller and thicker spatial distribution than younger (or relatively metal-rich/low-$\alpha$) populations. \\

This phenomenon constrains the structure formation process and can be well explained by an inside-out and upside-down scenario. It starts with a major infall of metal-poor gas. The gas with smaller angular momentum \textbf{first} cooled down and ignited the starburst formation process. The dense inner region caused a higher star-forming rate (SFR), so these populations are relatively more metal-poor but have higher $\alpha$-enhancement. The larger angular momentum gas subsequently inflowed and cooled down into the outer region. The outer gas may be enriched by the previous star formation and a lower SFR is required to occur the lower $\alpha$-enhancement. Meanwhile, it provides sufficient cooling time to form the larger but thinner-distributed stellar populations. This scenario confirms the model or simulation predictions. For instance, \cite{2018MNRAS.473..867K} performed N-body simulations and suggested that the thick disk was built-up in an inside-out and upside-down fashion, with older, smaller, and thicker populations being more metal-poor (see their Fig. 1). \\

Alternatively, kinematic heating is another effective mechanism that can thicken the older populations if they are formed from a relatively flat disk \citep{2017MNRAS.467.1154S}. \ed{ Actually, the observational evidence presented in this paper cannot clearly distinguish these two mechanisms. They may co-exist, which means the disk settles over time in an 'upside-down' formation scenario}, and the stellar populations are heated up after birth \citep{2021MNRAS.503.1815B}.\\

\ed{Now we can summarize the structure formation history of the entire MW disk as a hierarchically inside-out formation process.} The first stage is about the inside-out formation of the thick disk， which is claimed in this paper. The second stage is the consequently inside-out formation procedure of inner disk components, from the canonical thick (\hamp\,) disk to the \hamr\, disk and then to the \lamr\, disk (see HS22 for details). Finally, the third stage is the overall formation sequence from the thick (inner) disks to the thin (outer) disk shown in previous works (e.g., \citealt{2006ApJ...639..126B,2013ApJ...773...43B,2015ApJ...804L...9M}). However, it is worth noting that these similar structure formation patterns occur in opposite chemical radial gradients, indicating that they actually have different origins. The canonical thick disk formed earlier from highly turbulent gas. It quickly enriched the latter star formation region (with larger $L_Z$ or $R_g$) of the thick disk so that it caused an inverse [M/H] gradient. In contrast, the canonical thin disk formed mainly due to the second gas accretion of the MW so that the outer disk region was continually rejuvenated by lower- [M/H] gas resulting in the gradual decrease of the average stellar metallicity \citep{Lian20}.\\

\section{Conclusion}\label{sec:Conclusion}

Benefiting from the precise 6D kinematics provided by \ed{Gaia EDR3 and APOGEE DR17,} we can get the angular momentum and other orbital properties of the thick disk giant stars. Combining with the chemical abundances from the APOGEE and stellar ages derived from \cite{2018MNRAS.481.4093S}, we investigate the angular momentum of MAPs and their correlation with chemicals and stellar age. We summarize our main results as follows.\\

(1) The angular momentum components, $L_Z$ and $L_P$, of MAPs have significant variations across the [$\alpha$/M]-[M/H] plane that follow the chemical evolutionary trajectory of the thick disk stellar populations.  \\

(2) These variations are well quantified by the inverse $L_Z$ (or radial) gradient and the $L_P$ (or vertical) gradient of the metallicity ([M/H]), and the corresponding correlations between angular momentum components and the $\alpha$-enhancement ([$\alpha$/M]). The stars of metal-poor (or high-$\alpha$) MAPs have smaller and thicker distributions than those of metal-rich (or low-$\alpha$) MAPs.\\

(3) \textbf{The angular momentum components show significant age dependence}. It clearly suggests an inside-out structure formation scenario of the thick disk, which supports the infalling gas going through a gradually extending and cooling down process. \\

These results provide improved observational constraints on the chemical-spatial distribution of thick disk stars, which calls for rigorous and global chemical-evolution models leading to the angular momentum and chemical \textbf{correlations shown} in this paper. \\

We sincerely thank the anonymous referee for valuable comments and suggestion. This work is supported by  the National Natural Science Foundation of China (NSFC) under grants U2031139 and 12273091, the National Key R\&D Program of China No. 2019YFA0405501, and the science research grants from the China Manned Space Project with NO. CMS-CSST-2021-A08. This work has made use of data from the European Space Agency (ESA) mission Gaia (https://www.cosmos.esa.int/gaia), processed by the Gaia Data Processing and Analysis Consortium (DPAC; https://www.cosmos.esa.int/web/gaia/dpac/consortium). Funding for the DPAC has been provided by national institutions, in particular the institutions participating in the Gaia Multilateral Agreement. The stellar parameters, abundances, RVs from APOGEE DR17 were derived from the allStar files available at https://www.sdss.org/dr17/. Funding for the Sloan Digital Sky Survey IV has been provided by the Alfred P. Sloan Foundation, the U.S. Department of Energy Office of Science, and the Participating Institutions. 

%==============================================================
\software{Astropy \citep{2013A&A...558A..33A}, Numpy \citep{2011CSE....13b..22V}, Scipy \citep{2007CSE.....9c..10O}, Matplotlib \citep{2007CSE.....9...90H}.}

\bibliography{gradient}
\bibliographystyle{aasjournal}

\end{CJK*}	
\end{document}